\documentclass{aastex}
\usepackage{spr-astr-addons}
\shorttitle{Starburst-AGN connection}
\shortauthors{Gonz\'alez Delgado et al.}
\begin{document}


\title{The Starburst-AGN connection: The role of stellar clusters in AGNs \\
  }


\author{Gonz\'alez Delgado, R.M.\altaffilmark{1}, 
Mu\~noz Mar\'\i n, V.\altaffilmark{1},
P\'erez, E.\altaffilmark{1}}
\affil{Instituto de Astrof\'\i sica de Andaluc\'\i a (CSIC), Spain}

\author{Schmitt, H. \altaffilmark{2}}
\affil{Naval Research Laboratory, Washington, USA}

\author{Cid Fernandes, R.\altaffilmark{3}}
\affil{Universidade Federal de Santa Catarina, Brasil}


\begin{abstract}
Nuclear stellar clusters are a common phenomenon in spirals and in
starbursts galaxies, and they may be a natural consequence of the star
formation processes in the central regions of galaxies. HST UV imaging
of a few Seyfert 2 galaxies have resolved nuclear starbursts in
Seyfert 2 revealing stellar clusters as the main building blocks of the
extended emission. However, we do not know whether stellar clusters
are always associated with all types of nuclear activity. We
present NUV and optical images provided by HST to find out the role
that stellar clusters play in different types of AGNs (Seyferts and
LLAGNs). Also with these images, we study the circumnuclear dust
morphology as a probe of the circumnuclear environment of AGNs.
\end{abstract}


\section{Introduction}

\subsection{Starburst-AGN connection: The role of stellar clusters}

Nuclear stellar clusters are a common phenomenon in spirals, having
been detected in 50-70\% of these sources (Carollo et al.  1998, 2002;
B{\" o}eker et al. 2002, 2004). Therefore stellar clusters are a natural
consequence of the star formation processes in the central region of
spirals. On the other hand, evidence has been accumulating during the
past few years about the ubiquity of black hole (BH) in the nuclei of
galaxies. Furthermore, the tight correlation of the BH mass and
stellar velocity dispersion (Ferrarese \& Merrit 2000; Gebhardt et
al.\ 2000) implies that the creation and evolution of a BH is
intimately connected to that of the galaxy bulge. Recently, in a
HST survey in the Virgo Cluster, C\^ot\'e et al. (2006) have detected
compact sources in a comparable fraction of elliptical galaxies.
These compact stellar clusters, referred to as nuclei by the
authors (see also Ferrarese et al 2006a), have masses that scale
directly to the galaxy mass, in the same way as do the BH masses in
high luminosity galaxies (Ferrarese et al. 2006b). Therefore, a
natural consequence of the physical processes that formed present-day
galaxies should be the creation of a compact massive object in the
nucleus, either a BH and/or a massive stellar cluster.

\subsection{Starburst-AGN connection: Seyfert galaxies}

According to the unified scheme of AGNs, the main components of a Seyfert nucleus are: 1) A BH and its associated accretion disk. 2) A circumnuclear dusty torus that collimate the AGN radiation through its polar axis. So, a Seyfert 2 nucleus should be a Seyfert 1 that is viewed close to the equatorial plane.  3) A mirror of dust and warm electrons located along the polar axis of the torus, that reflects and polarizes the AGN radiation. Seyfert 2 nuclei exhibit a featureless continuum (FC) that comprises much of the near-UV. It was long-thought that this FC was light from the hidden Seyfert 1 nucleus. Cid Fernandes \& Terlevich (1995) proposed that a heavily-reddened starburst provides this FC.

Because of the high sensitivity of UV wavelengths to the presence of massive stars, HST UV observations were obtained for a few Seyfert 2 galaxies to probe the role of starbursts in their nuclei and the origin of the FC (Heckman et al. 1997; Gonz\'alez Delgado et al. 1998). HST high spatial resolution imaging shows that the UV continuum source is spatially extended ($\sim$ 100 pc) and it is resolved in knots with sizes of a few parcsecs and properties similar to the stellar clusters detected in starburst galaxies. UV  spectra for four galaxies, corresponding to the central $\sim$ 100 pc, were obtained.  The data provided direct evidence of the existence of a nuclear starburst that accounts for the FC. Absorption features formed in the photospheres and in the stellar winds of massive stars are detected. Their UV colors indicate that the starburst is quite reddened. Their bolometric luminosities are similar to the estimated luminosities of the hidden Seyfert 1 nuclei. 
Subsequently, near-UV and optical spectra of a large sample of Seyfert 2 were obtained proving unambiguous identification of circumnuclear starburts in $\sim$40$\%$ of nearby Seyfert 2 galaxies as well as their energetic significance (Gonz\'alez Delgado et al. 2001; Cid Fernandes et al. 2001). These results have been latter confirmed  studying a large sample observed by SDSS (Heckman et al. 2004; Wild et al. 2007).

\subsection{Starburst-AGN connection: LLAGNs}

Low-luminosity active galactic nuclei (LLAGNs) constitute a significant  fraction of the nearby AGN population. These include LINERs, and transition-type objects (TOs, also called weak--[OI] LINERs) whose properties are in between classical LINERs and H{\sc ii} nuclei.  LLAGNs comprise $\sim$ 30$\%$ of all bright galaxies and are the most common type of AGN (Ho, Filippenko \& Sargent 1997; hereafter HFS97).  What powers them and how they fit in the global picture of AGN has been at the forefront of AGN research for over two decades. Are they all truly ``dwarf Seyfert galaxies'' powered by accretion onto a nearly dormant super-massive black hole, or can some of them be explained at least partly in terms of stellar processes? 

HST observations at UV wavelengths of a few LLAGNs have also proven that at least weak-[OI] LINERs could be powered by young massive stars (Maoz et al. 1998; Colina et al. 2002). Nuclear stellar clusters are detected in these objects through high spatial resolution UV imaging and spectra. NGC 4303 is probably the best example (Colina et al. 2002).  Additional evidence comes from optical studies, in which we have focused on the study of the stellar population in the nuclei and circumnuclear region of LLAGNs (Cid Fernandes et al.\ 2004; Gonz\'alez Delgado et al.\ 2004). These studies
identified a class of objects, called ``Young-TOs'', which are clearly
separated from LINERs in terms of the properties and spatial
distribution of the stellar populations. They have stronger stellar
population gradients, a luminous intermediate age stellar population
concentrated toward the nucleus ($\sim$100~pc) and much larger amounts
of extinction than LINERs. These objects, which underwent a powerful
star formation event $\sim$ 1 Gyr ago, could correspond to
post-Starburst nuclei or to evolved counterparts of the Seyfert 2 with
a composite nucleus, characterized too by harboring nuclear starbursts. 

However, HST UV monitoring observations of 17 LLAGNs have detected variability with amplitudes from a few percent to 50$\%$ (Maoz et al. 2005). The variability is more frequently detected in those LINERs that have a compact radio core, as expected from bona fide AGNs. Thus, it is not clear which fraction of LLAGNs have a nuclear stellar cluster and which are "dwarf Seyfert galaxies" like those that show UV variability.

\subsection{Goals of the project}

The main goals of the project are to detect and characterize the properties of nuclear and circumnuclear  compact sources in different types of AGNs (type 1 and 2 Seyferts, LINERs and TOs), to  determine their nature and their evolution. 
The high angular resolution provided by HST is crucial to determine the morphological structure of the nuclear and circumnuclear region of these objects.
Due to the steep increase of the surface brightness towards the center of early type galaxies, the near-UV is the ideal wavelength range to isolate the contribution of the nuclear and circumnuclear sources from the bulge component (see Figure 1). Although young
stellar clusters are powered by massive stars (O and B), which emit
most of their flux at the far-UV, the near-UV  can also trace equally
well their distribution (see Figure 1).  In addition, near-UV is more
useful than far-UV observations to detect for example clusters with intermediate
ages (few 100 Myr), considering the weakness of these populations at
shorter wavelengths. Further information at longer wavelengths (e.g. optical and NIR) is also very useful to get colors 
and to estimate the stellar clusters ages. Additionally high spatial resolution optical images are essential to trace the morphology of the circumnuclear dust which is a valuable probe of the presence of cold interstellar gas in galaxies, and it is very sensitive to the perturbations that drive the gas toward the center and feed the AGN. Thus, the circumnuclear dust can also provide
relevant information about the origin of nuclear activity. 

To reach these goals we are carrying out several projects with HST+ACS
imaging at the near-UV wavelengths a sample of Seyferts (ID. 9379, PI. Schmitt) and LLAGNs
(ID. 10548, PI.  Gonz\'alez Delgado). These observations are
complemented with WFPC2 optical data retrieved from the HST archive. 
Here we present a summary of the the first results obtained for the sample of Seyferts and LLAGN galaxies. 
Further explanations are in Mu\~noz-Mar\'\i n et al (2007) and Gonz\'alez Delgado et al (2007). 

\section{Results: Seyfert galaxies}

The sample contains 75 Seyferts that were observed with HST+ ACS/HRC (F330W). These observations provide a pixel scale of 0.027 arcsec. The filter F330W has a bandwidth of $\sim400$ \AA \,centred around 3300 \AA, therefore the only strong emission lines contributing to this filter are [NeV]$\lambda$$\lambda$3346,3426. This type of emission will be normally extended, thus is not a problem for measuring compact objects such as clusters.

The 63$\%$ of these objects are classified as 
Seyfert 2 (49 objects of the total), 19$\%$ as intermediate types Sy1.8-1.9 (14 objects), and 19$\%$ as Seyfert 1 (14 objects). The sample was built from the list of Seyferts that were previously observed by HST at optical and NIR wavelengths. 
  In order to understand the possible biases of this sample we have compared the general properties of these galaxies with those in two bona-fide samples of Seyfert galaxies in the literature, the CfA and RSA Seyfert subsamples. From the 48 Seyfert galaxies in the CfA catalogue presented in Huchra \& Burg (1992), 24 are in our sample as well. On the other hand, 38 out of 75 of our galaxies belong to the extended RSA Seyfert sample compiled by Maiolino \& Rieke (1995). We thus explore 50\% of CfA and 42\% of RSA, with only 10 of our galaxies occurring in both of them. In terms of distances, morphological types, and axial ratios, our sample follows similar distribution to the RSA and CfA samples.
  
Figure 2 shows four examples of the Seyfert galaxies observed.
These images show the distribution of circumnuclear stellar clusters that formed at spatial scale of several 100 pc to 1 kpc from the galaxy center. Compact nuclear sources are detected in all the Seyfert 1 galaxies, but the nucleus is obscured in  some of the Seyfert 2 (e.g. NGC 1672). In general, the NUV emission morphology of these objects is  as irregular as varied. There are many different features within the sample: star-forming rings, spirals, clumpy diffuse light emission, plain PSF-dominated objects, complete lack of compact nucleus, etc. Seyfert 1 possess a bright star-like nucleus, which precludes the observation of the inner morphology in many of the cases. In several galaxies, regions of star-formation and rings can be seen in the images as well. The morphology of intermediate type Seyferts is more varied. Some objects have a compact nucleus. The morphology can be clumpy or diffuse, and some objects show dust absorption features or ionization cones. For the Seyfert 2  galaxies the morphology is mostly clumpy, with frequent star-formation regions. These are often arranged in rings or spiral arms.  There are some objects showing instead a biconical or symmetrical structure as ionization cones. 

We have carried out a general analysis of the near-UV images of this sample, consisting in the identification of unresolved compact nuclear sources, extraction of surface brightness profiles, photometry, determination of compactness and asymmetry parameters and identification of the star cluster population.  We have compared the results of the analysis among different Seyfert types. We have also estimated the fraction of the NUV light that is provided by the circumnuclear stellar clusters. We have found that star clusters are more often seen in Seyfert 2 than Seyfert 1, being much more important in Seyfert 2 than in other Seyfert types the contribution of the clusters to the total flux.

\section{Results: LLAGNs}

The raw sample of this study contains LLAGNs from the Palomar catalogue (HFS97) for which we have already studied the circumnuclear stellar populations via STIS and ground based optical spectra (Cid Fernandes et al 2004; Gonz\'alez Delgado et al 2004), and there are WFPC2 optical images available in the HST archive. Following the HFS97 emission-line classification, our subset contains
32 LINERs ([OI]/H$\alpha > 0.17$ ) and 25 TOs ([OI]/H$\alpha \le
0.17$), i.e., 34 and 38\% of whole HFS97 LLAGN sample, respectively.
In our previous studies we proposed a slight modification of this criterion,
setting the LINER/TO dividing line at [OI]/H$\alpha = 0.25$. According
to this definition, to be used throughout the rest of this paper, our
sample includes 36\% of the TOs and 34\% of the LINERs in the HFS97
catalog.

This sample has already been observed with HST+ACS with HRC (0.027arcs/pix) and through the F330W, as the Seyfert sample. The analysis of this data is in progress, and the results will be presented soon. Here, we report the results obtained from the analysis of the archive WFPC2 images obtained through any of these optical filters: F555W, F547W, F606W, F791 and F814W. Most of the objects were imaged with the nucleus at the PC, thus with a spatial sampling of 0.046 arcsec/pixel. The mean spatial sampling  of these images is 10 pc. 
With these data we have built an atlas that includes structural
maps for all the galaxies, useful to identify compact nuclear
sources and, additionally, to characterize the circumnuclear
environment of LLAGNs, determining the frequency of dust and its
morphology.  The structural maps are obtained through two ways. 1) By the unsharp masking technique. The unsharp masked image is obtained by dividing the original frame by a smoothed version obtained with a
31$\times$31 median boxcar kernel. Figure 3 shows several examples. 2) By fitting the central light distribution by elliptical isophotes. The structural map is obtained subtracting the model fit from the original image (see Figure 4). Nuclear compact sources can be also identified through the surface brightness analysis. These sources are identified sometimes for rising above the
inward extrapolated surface brightness cusp at small radii. 

The most remarkable result is that dust is almost ubiquitous in
LLAGNs. Only 12\% of the galaxies of our sample do not display dust
features.The morphology is quite diverse, from nuclear disks, to filaments and
lanes chaotically distributed, to well organized nuclear spiral
arms.  Dust is as frequent as in Seyferts (Martini et al. 2003), but
the distribution of the dust morphology is different. Most of the 
Seyferts have dust spirals or dust filaments and
arcs chaotically distributed; the dust disk morphology and non-dusty
class are quite rare, while LLAGNs display dust disk morphology. 
Dividing the LLAGN sample in TOs and LINERs, we have found that 
chaotic filaments are as frequent as dust spirals in both types of LLAGNs; 
but nuclear disks are mainly seen in LINERs.  
These results suggest an evolutionary sequence of the dust in LLAGNs, LINERs
being the more evolved systems and Young-TOs the youngest.

From the central  photometry measurements and surface brightness profiles, we have found 
that LINERs and TOs have both similar central magnitude and
surface brightness, but LLAGNs with young and intermediate age
populations are brighter than Old-TOs and LINERs. We have not found
any correlation between the presence of nuclear compact sources and
the emission line spectral type, ie., LINERs are as frequently
nucleated as TOs.  However, the centers of Young-TOs are brighter than
the centers of Old-TOs and LINERs. The difference in magnitude and
surface brightness can be even larger if we account for internal
extinction, since Young-TOs are dustier. This result indicates that 
Young-TOs are separated from other type of LLAGNs also in terms 
of their central brightness, in addition of the properties and spatial 
distribution of the stellar population. In other to further investigate the 
nature of the nuclear components, we have compared the distribution of the nuclear radio and optical 
luminosities of our LLAGNs with those of FRI radio galaxies and Seyfert galaxies (see Chaiberge, Capetti \& Macchetto 2005).
We found that the LINERs of our sample with compact source follow the same relation that AGNs, but young TOs are shifted from the correlation, indicating indirectly a stellar origin for the nuclear compact sources in these objects.

\section{Conclusions}

The main conclusions obtained for Seyfert sample are: 1) The UV morphology is very irregular, with 
clumpy and compact structures in most of the cases. 2) Seyfert  1 are very compact and completely PSF dominated; 
while Seyfert  2 have the nucleus resolved. 3) Circumnuclear stellar clusters  are more common in Seyfert  2 than in Seyfert 1; but the distribution of the flux in stellar clusters does not change much among different Seyfert types, when only galaxies with clusters detected are considered. 4) The contribution of the clusters to the total flux is more important in Seyfert  2 than in other Seyfert types. 

The main results obtained for LLAGNs  are: 1) We have not found any
correlation between the presence of nuclear compact sources and
emission-line type.  Thus, nucleated LINERs are as frequent as
nucleated TOs.  2) The nuclei of "Young-TOs" are brighter than the
nuclei of "Old-TOs" and LINERs. These results confirm our previous
results that Young-TOs are separated from other LLAGNs classes in
terms of their central stellar population properties and brightness.
3) Circumnuclear dust is detected in 88$\%$ of the LLAGNs, being
almost ubiquitous in TOs.  4) The dust morphology is complex and
varied, from nuclear spiral lanes to chaotic filaments and nuclear
disk-like structures. Chaotic filaments are as frequent as dust
spirals; but nuclear disks are mainly seen in LINERs.  These results
suggest an evolutionary sequence of the dust in LLAGNs, LINERs
being the more evolved systems and Young-TOs the youngest.

\acknowledgments

This work has been funded with support from the Spanish Ministerio de
Educaci\'on y Ciencia through the grants AYA2004-02703 and
AYA2007-64712, and co-financed with FEDER funds, and by the NASA grants HST-GO-9379.01-A and HST-GO-10548.01-A. Basic research in astronomy at NRL is supported by 6.1 base funding. 
The data used in this work come from observations made
with NASA/ESA Hubble Space Telescope, and some of them are obtained from the STScI data
archive.


\begin{figure}
\begin{center}
  \plotone{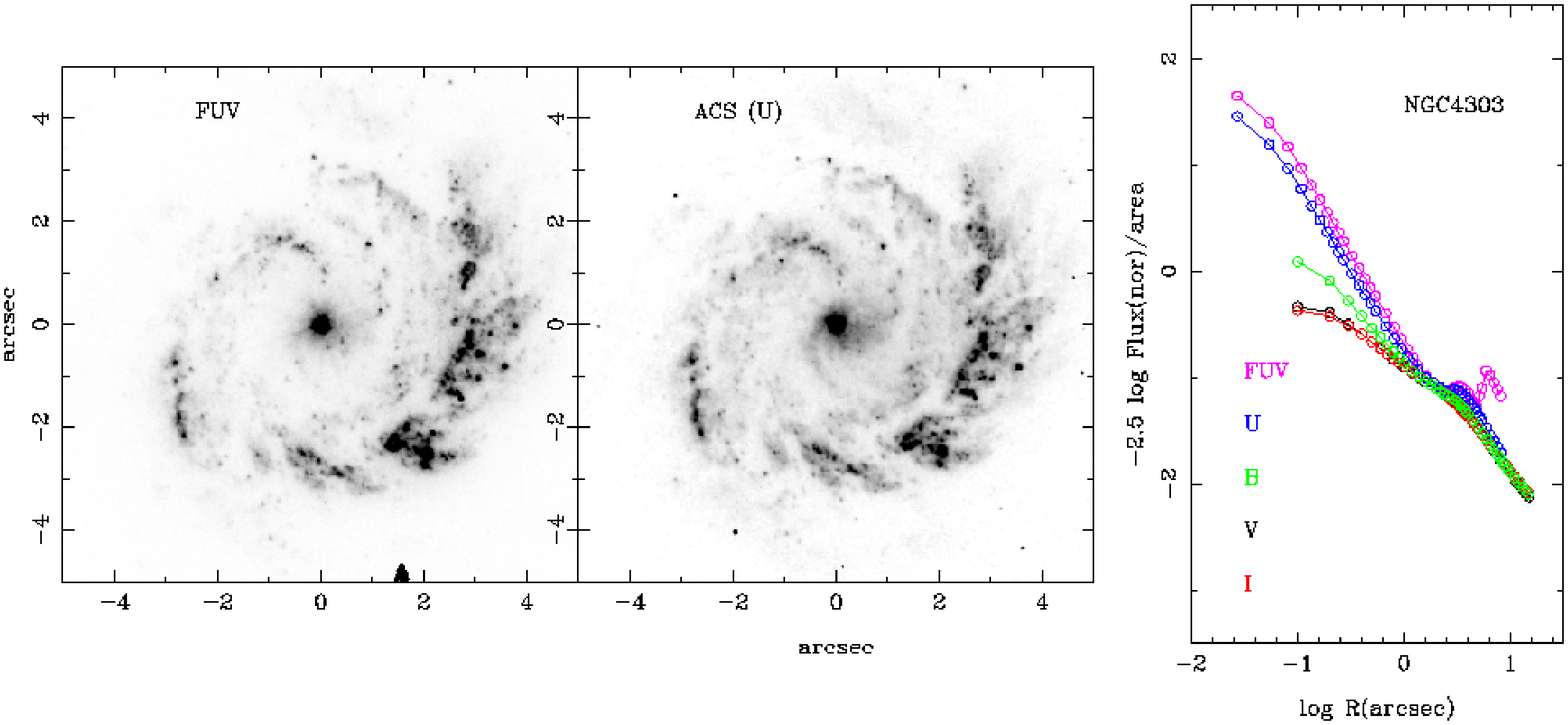}
  \plotone{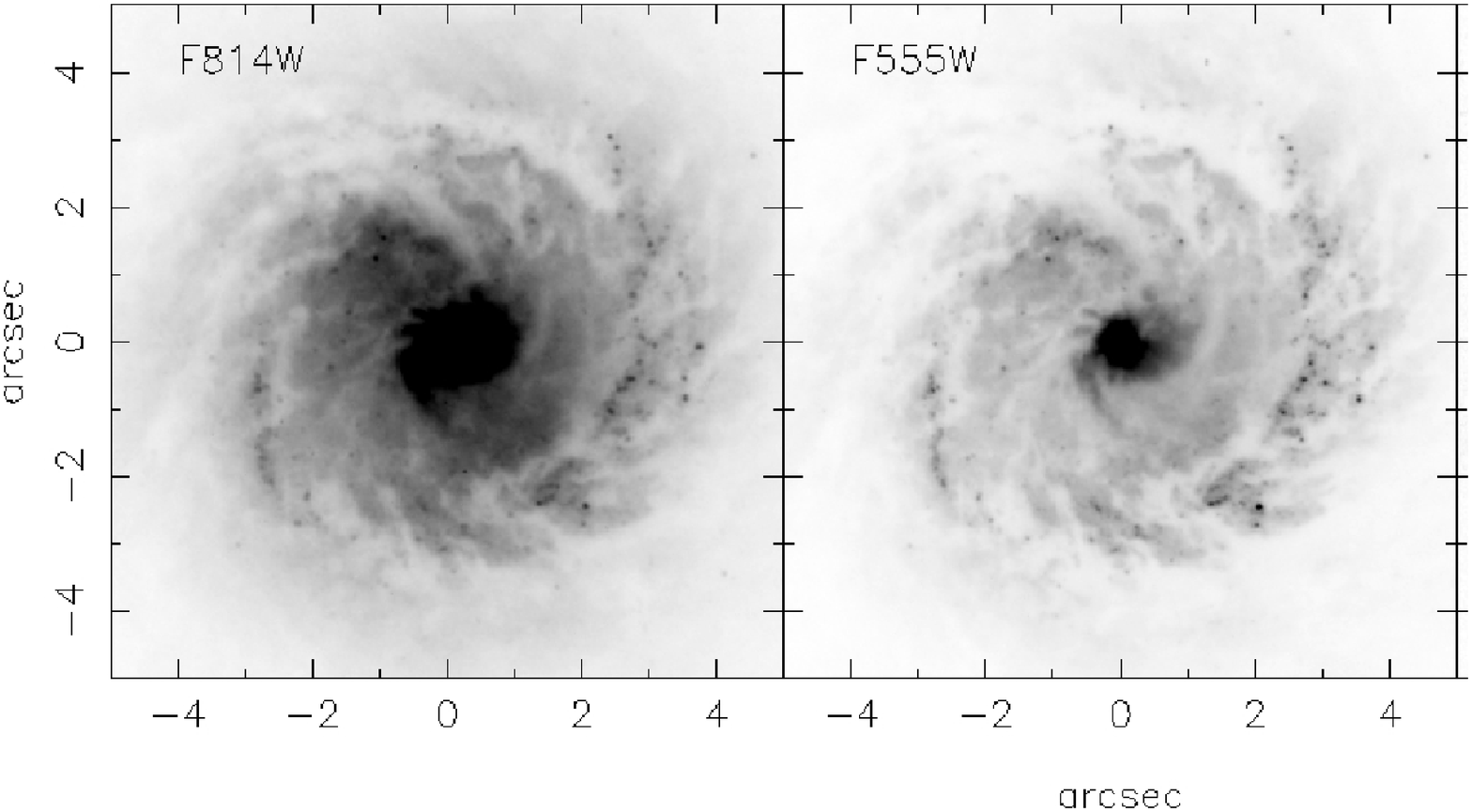}
 \end{center}    
\caption{ HST (ACS) images of the LLAGN NGC4303. The top pannel (from left to right) shows the FUV, NUV images and the surface brightness profiles for the FUV, NUV, and optical filters. The lower pannel show the images through the F814W and F555W filters.  These figures show a more significant detection of the nuclear source at near-UV wavelenght, and a similar fux distribution at the far-UV and near-UV.}
\label{fig_1}   
\end{figure}

\begin{figure}
\begin{center}
\plottwo{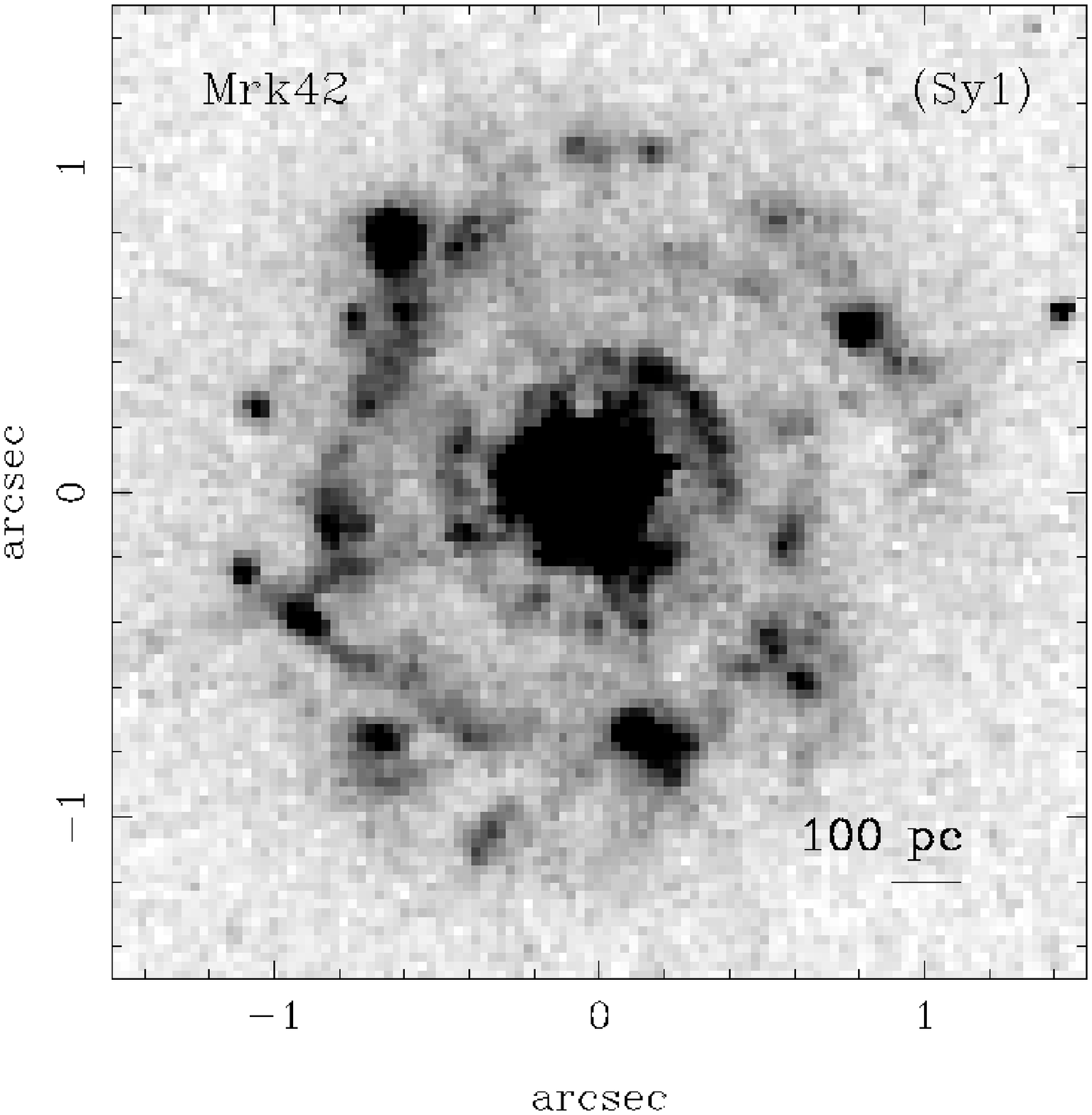}{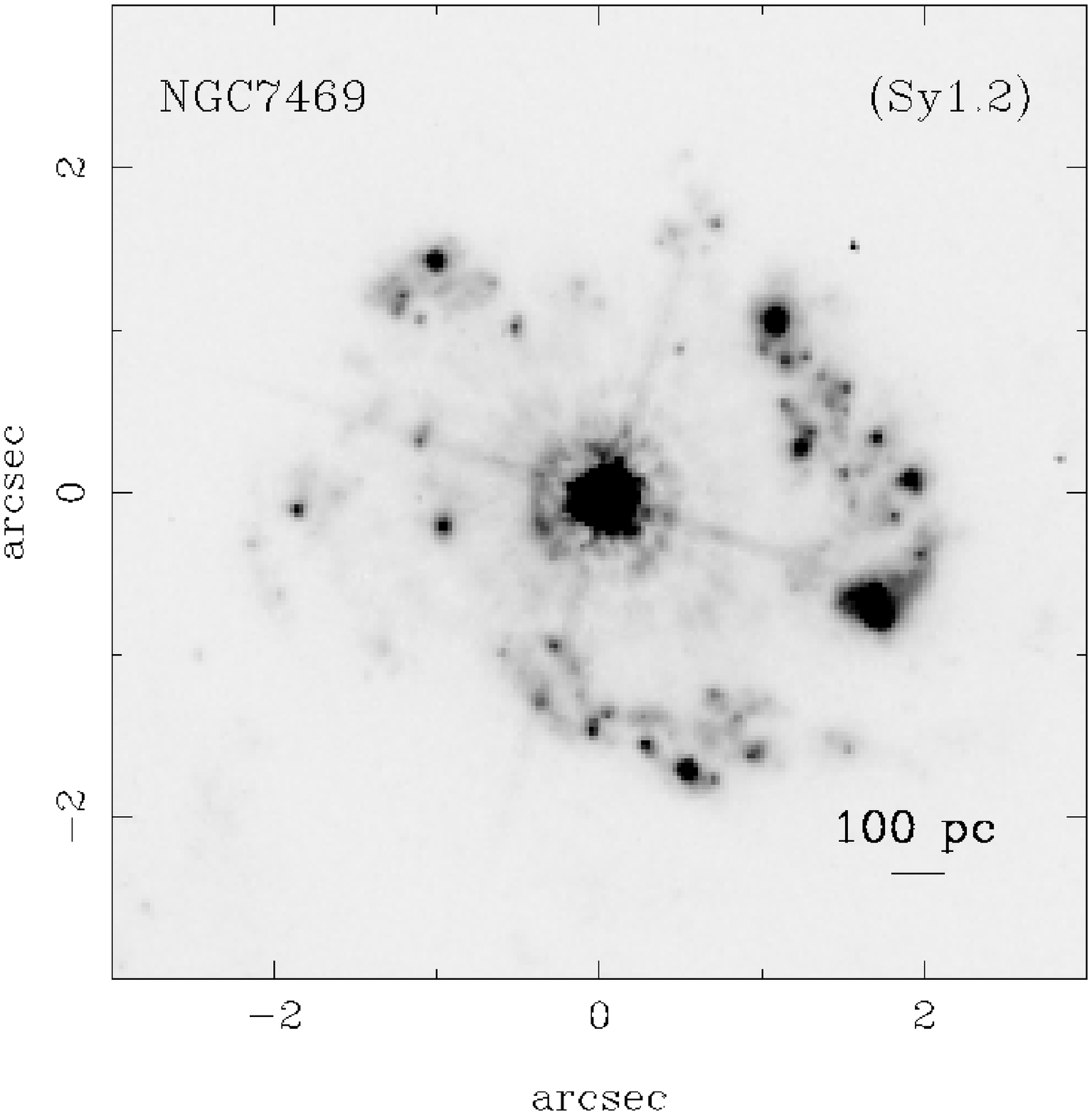}
\plottwo{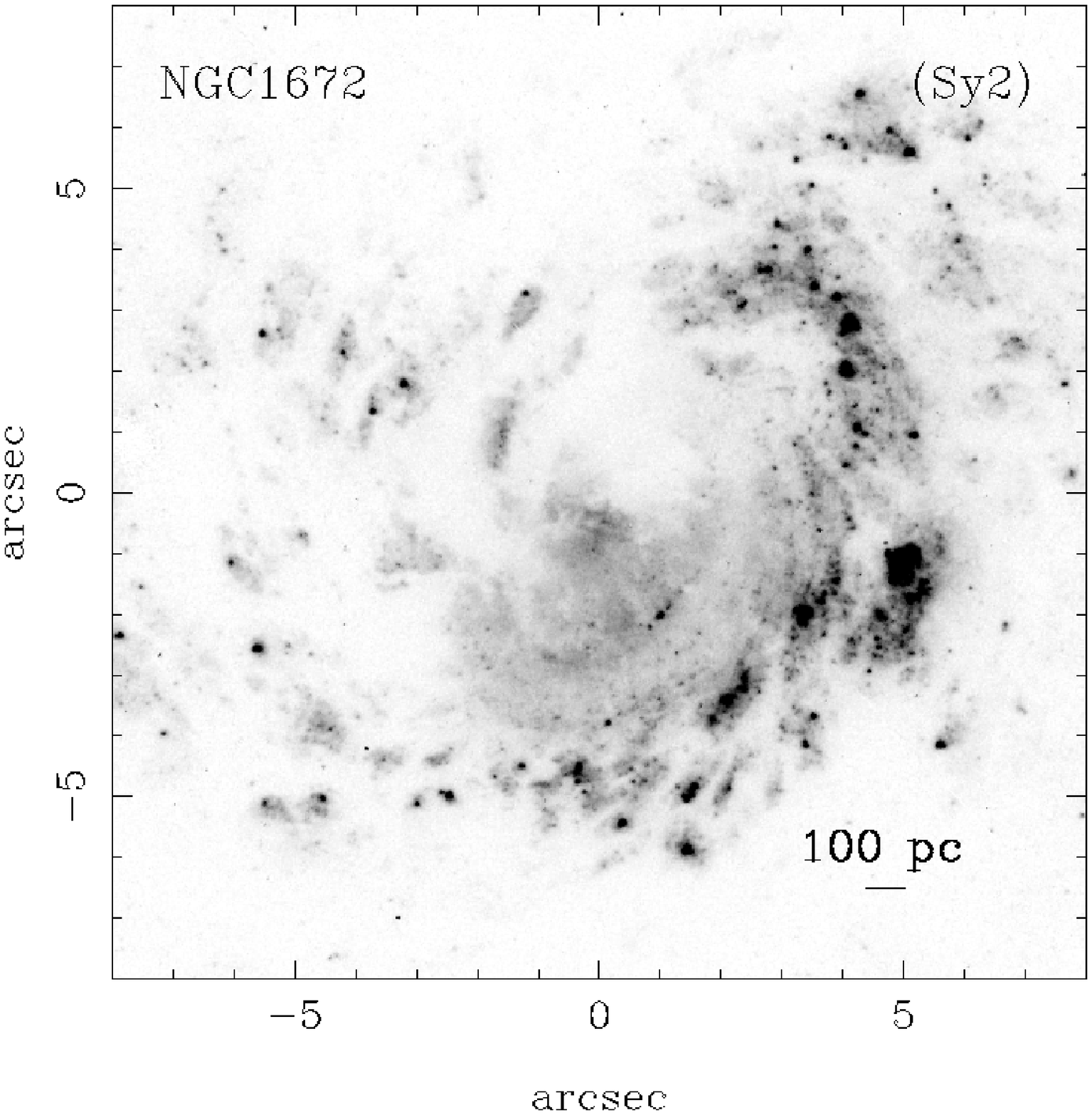}{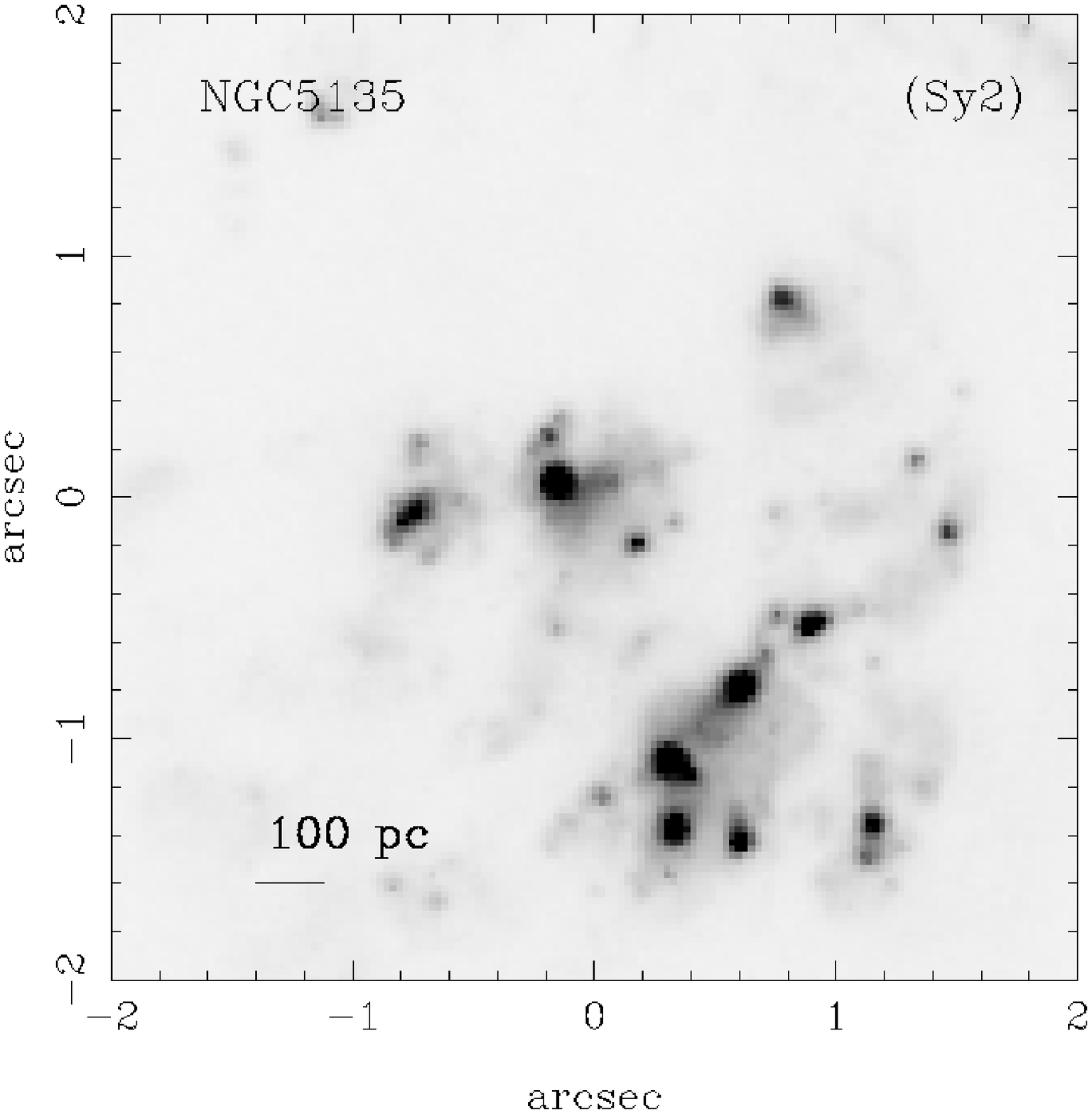}
 \end{center}    
\caption{HST+ACS (F330W) images of four examples of Seyfert galaxies, NGC 1672 (Sy2), Mrk 42 (Sy1), NGC 5135 (Sy2) and NGC 7469 (Sy1.2), that show circumnuclear stellar clusters. }
\label{fig_2}   
\end{figure}

\begin{figure}
\begin{center}
\plottwo{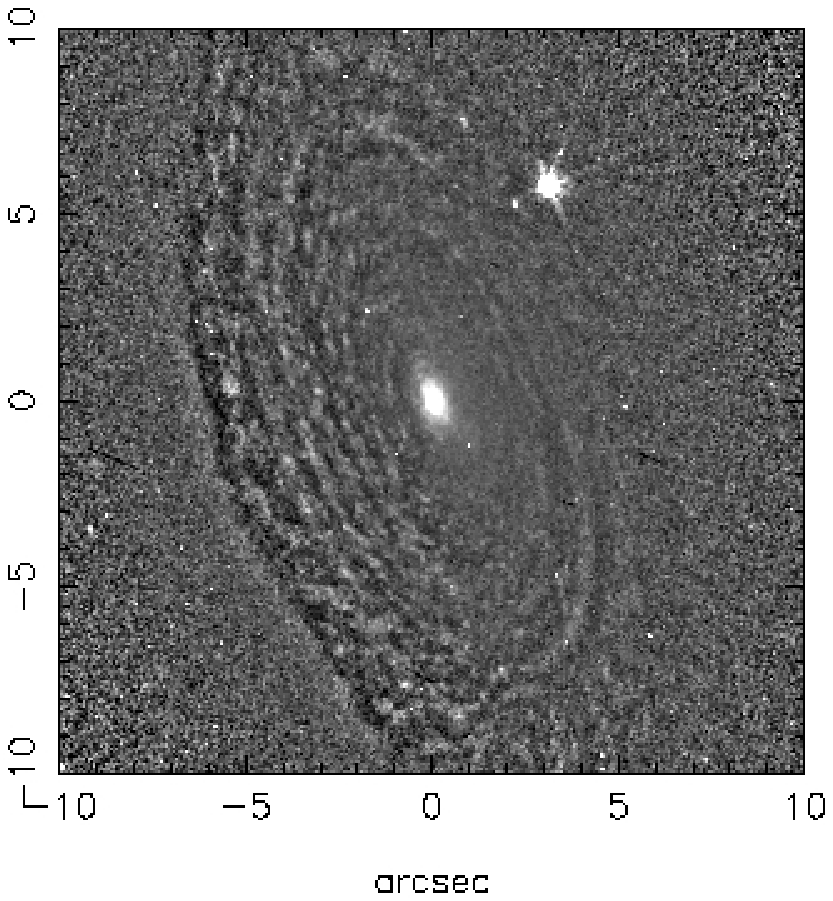}{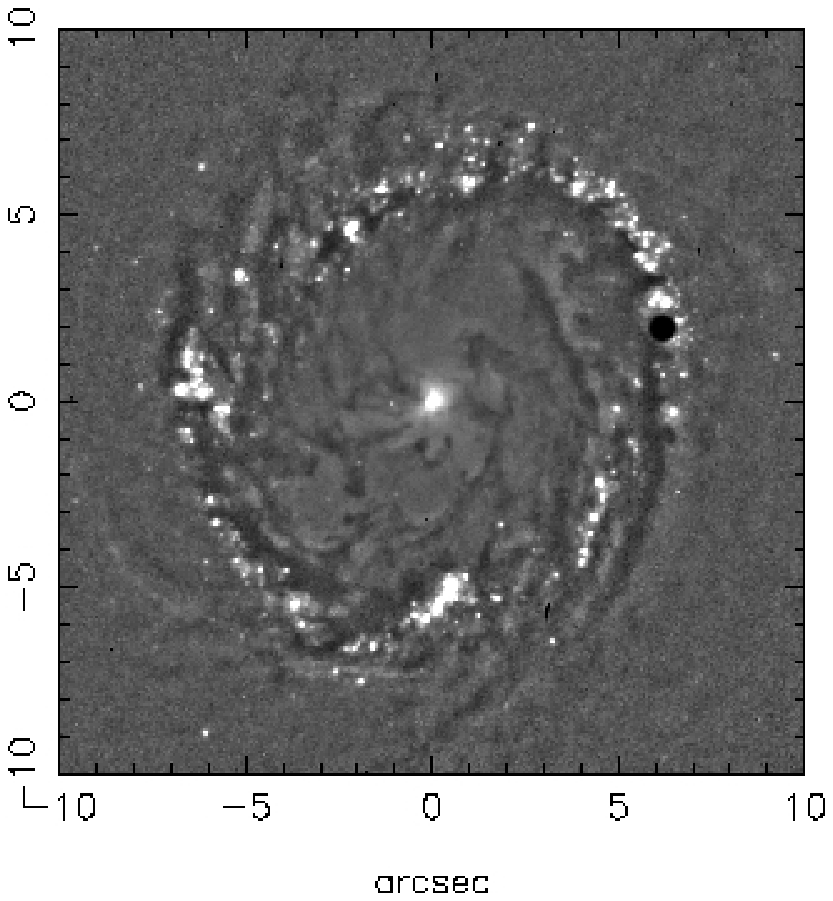}
\plottwo{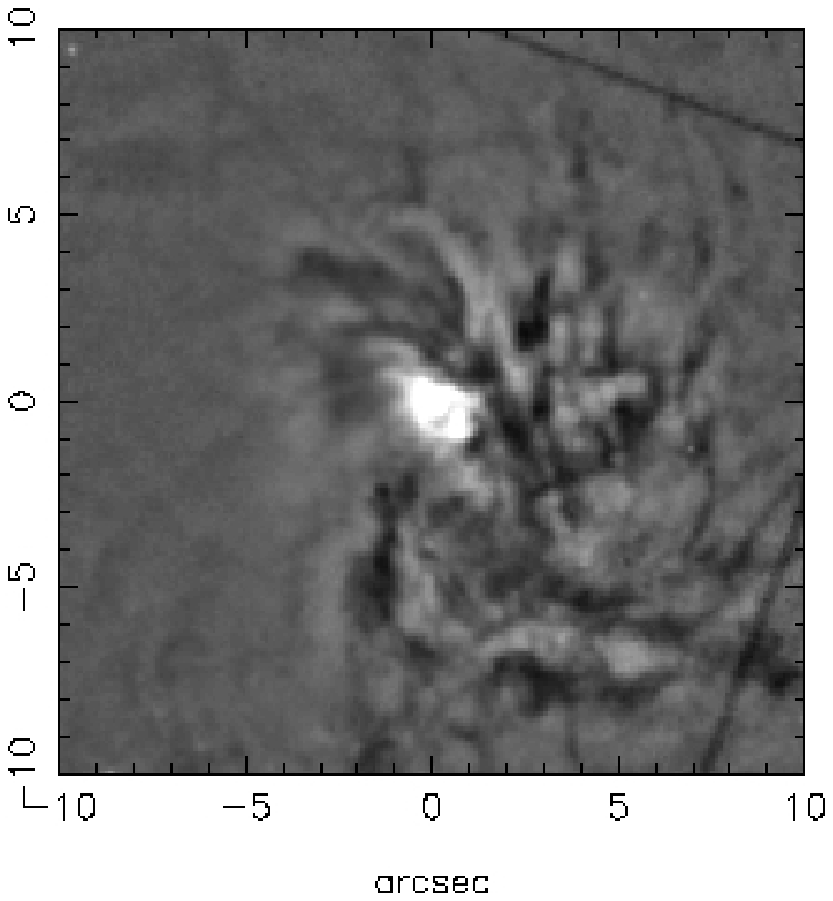}{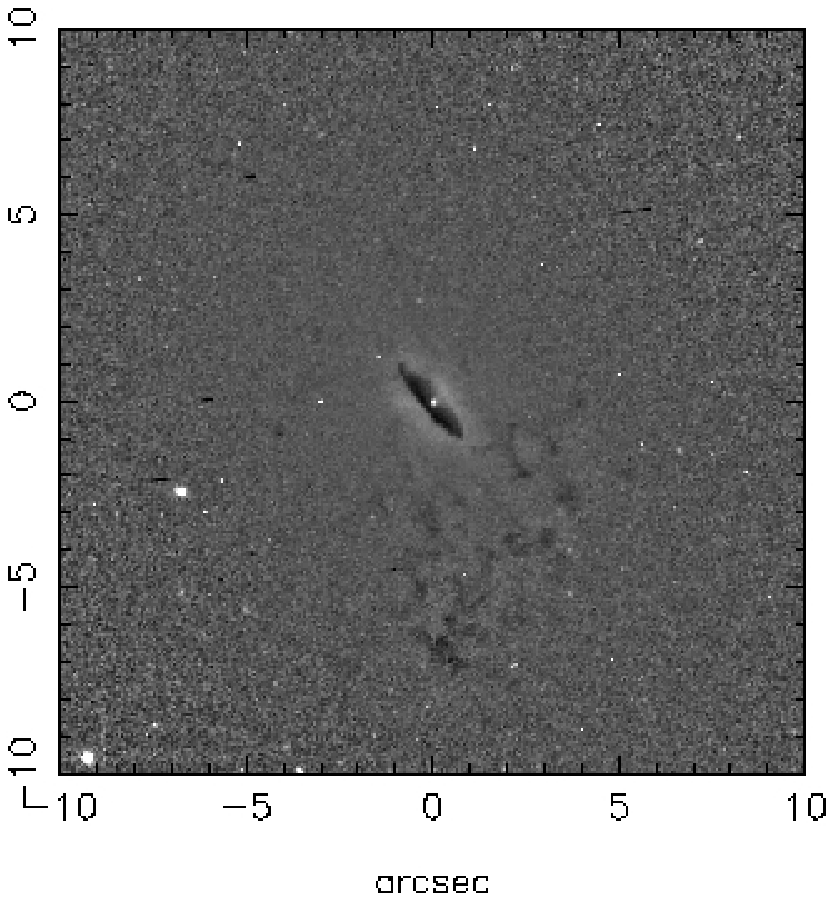}
 \end{center}    
\caption{Four examples of the circumnuclear dust morphology in LLAGNs. At the top panels: NGC 1161 (left) and NGC 4314 (right), two TOs nuclear spirals dust morphology. At the lower panels: NGC 3368 (left) and NGC 315 (right), a TO and a LINER with chaotic circumnuclear dust and disks nuclear ring, respectively. }
\label{fig_3}   
\end{figure}

\begin{figure}
\begin{center}
\plotone{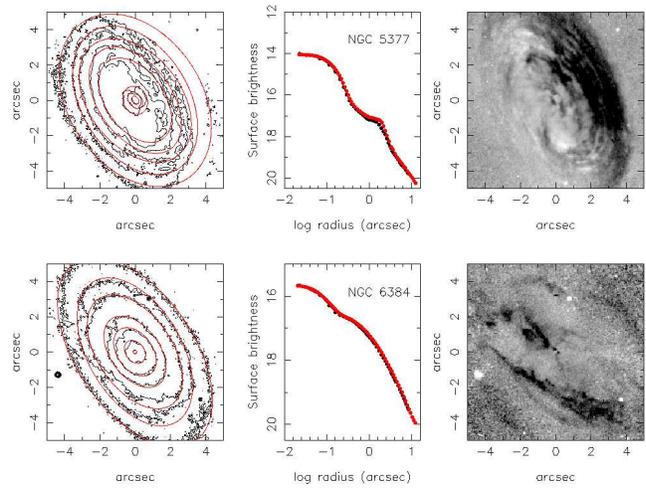}
  \end{center}    
\caption{The isophotal analysis (left panel) done for NGC 5377 (upper panel) and NGC 2787 (lower panel). The center panels show the surface brightness profiles and the right panel the dust maps. }
\label{fig_1}   
\end{figure}

\end{document}